\documentclass[aps,pre,twocolumn,showpacs,amsmath,amssymb]{revtex4}
\usepackage{bm}
\usepackage{graphics}

\begin{document}
\draft

\newcommand{\pp}[1]{\phantom{#1}}
\newcommand{\be}{\begin{eqnarray}}
\newcommand{\ee}{\end{eqnarray}}
\newcommand{\ve}{\varepsilon}
\newcommand{\vs}{\varsigma}
\newcommand{\Tr}{{\,\rm Tr\,}}
\newcommand{\pol}{\frac{1}{2}}
\newcommand{\ba}{\begin{array}}
\newcommand{\ea}{\end{array}}
\newcommand{\bear}{\begin{eqnarray}}
\newcommand{\eear}{\end{eqnarray}}
\title{
Quantum Circa-Rhythms
}
\author{Diederik Aerts$^1$, Marek Czachor$^2$, and  Monika Syty$^2$}

\address{
$^1$ Centrum Leo Apostel (CLEA) and Foundations of the Exact Sciences (FUND)\\
Brussels Free University, 1050 Brussels, Belgium\\
$^2$ Katedra Fizyki Teoretycznej i Metod Matematycznych\\
Politechnika Gda\'nska, 80-952 Gda\'nsk, Poland}

\begin{abstract}
A class of solutions of nonlinear von Neumann equations exhibits rhythmic properties analogous to those found for nonlinear kinetic equations employed in phenomenological description of biochemical oscillations. As opposed to kinetic equations, typically solved numerically and involving Hill-type nonlinearities, our exact solutions are derived by means of soliton techniques and involve a simple quadratic nonlinearity.
\end{abstract}
\pacs{87.23.Cc, 03.67.Mn, 05.45.Yv}
\maketitle

\section{Introduction}

Perhaps the greatest achievement of a pre-quantum phenomenology in chemistry was the Mendeleev periodic table of elements. The table, in particular, allowed to predict new elements with correctly anticipated properties. But to {\it really\/} understand the structure of  the table one had to invent a simply-looking equation 
\be
i|\dot \psi\rangle=H|\psi\rangle\label{S}
\ee
where $|\psi\rangle$ is a vector from a Hilbert space and $H$ a self-adjoint operator.
 
The Schr\"odinger equation (\ref{S}) with appropriately chosen $H$ allowed to understand the origin and types of chemical bonds, that is, the deep structures behind the Mendeleev table. Although the identification of $H$ for a concrete molecule was of crucial importance, there were certain features of the abstract equation that allowed to determine general properties of those systems whose dynamics could be governed by (\ref{S}). For example, a self-adjoint $H$ was related to conservative and isolated sytems. The conserved quantity is the average energy 
$E=\langle \psi|H|\psi\rangle$. Conservative systems that are non-isolated (i.e. non-trivially interact with the outside world) were described by another simply-looking equation 
\be
i\dot\rho=[H,\rho]\label{vN}
\ee
where $\rho$ is a trace-class positive operator acting in the same Hilbert space and $H$ is the same self-adjoint operator. Eq.~(\ref{vN}) is known as the von Neumann equation. (\ref{vN}) is equivalent to (\ref{S}) if $\rho$ projects on the direction of $|\psi\rangle$. The invariant 
$\langle \psi|H|\psi\rangle$ is replaced by $E=\Tr (H\rho)$. If $|j\rangle$ is a normalized vector, then the diagonal element $\rho_{jj}=\langle j|\rho|j\rangle=\Tr P_j\rho=p_j$ is a probability associated with the proposition represented by $P_j=|j\rangle\langle j|$. The probability is time dependent if $[H,P_j]\neq 0$.  

Thinking of the present-day chemistry or biochemistry one is struck by the ubiquity of nonlinear kinetic equations \cite{Prigogine,Goldbeter}. Physically, all kinetic equations describe dynamics of probabilities and in this sense are close to (\ref{vN}). Typical kinetic equations describe open systems and the associated probabilities are related with commuting propositions. Nonlinearity occurs whenever there are feedbacks or cooperative phenomena, and details of evolution depend on the number of substrates and phenomenology used to model right-hand-sides of the equations. 

Similarly to the catastrophe theory, phenomenological modelling is to some extent an art 
\cite{Thom}. Having experience with nonlinear evolutions one is capable of choosing the right-hand-sides of the equations in such a way that certain qualitative elements of the dynamics will have to occur, and then it remains to play with parameters. 
Not trying to diminish the role of phenomenological modelling we have to admit that phenomenology is a two-edged sword. On the one hand, it links qualitative analysis with quantitative predictions, and often reveals aspects that are initially unexpected. On the other hand, however, shapes of the equations are dictated by our needs.  
This should be contrasted with (\ref{S}) and (\ref{vN}) which can be derived from first principles.

A general property of systems described by (\ref{S}) or (\ref{vN}) is the 
{\it exact\/} linearity of the evolutions they generate. Linear evolutions practically do not occur in biological and chemical systems, so linearity is in this context a drawback. Sometimes a linear approximation is justifed if some paramters are small, but the generic case is nonlinear. 
In Ref. \cite{I} we have argued that a candidate equation describing a  conservative quantum system in a feedback with some environment is the nonlinear generalization of the von Neumann equation 
\be
i\dot\rho=[H,f(\rho)]\label{fvN}
\ee
where $f(\rho)$ is an operator function satisfying the 
``no-feedback, no-nonlinearity" property, formally meaning that 
$f(\rho)=\rho$ if $\rho$ is a projector on some direction $|\psi\rangle$ in a Hilbert space. Eq.~(\ref{fvN}) is integrable in the sense of soliton theory for any $f$. The time-invariant is now $E=\Tr \big(Hf(\rho)\big)$. A soliton technique of solving (\ref{fvN}), based on Darboux transformations,  was introduced in \cite{LC}, further developed in \cite{UCKL}, and recently generalized in \cite{CCU}. 
Therefore, as opposed to the kinetic equations that typically have to be solved numerically 
(for exceptions cf. \cite{num1,num2,num3}), we can work with exact analytic solutions. This allows us to have access to the time scales and precision numerically  unavailable. 

The set of solutions of (\ref{fvN}) contains all the solutions of form $\rho=|\psi\rangle\langle\psi|$ where $|\psi\rangle$ satisfies (\ref{S}). A new class of ``self-switching" solutions occuring only for nonlinear $f$ was discovered in \cite{LC} and applied in \cite{I} in the context of morphogenesis. These solutions are characterized by three regimes involving a single switching between two asymptotically linear evolutions. 

Still, looking for the most striking manifestations of biochemical nonlinear evolutions one arrives at phenomena that are {\it rhythmic\/}. Periodic switching, with periods ranging from fraction of a second to years, is encountered at all levels of biological organization, and its successful modelling in terms of nonlinear kinetic equations is one of the great achievements of computational biology (for a review cf. \cite{Goldbeter2002}). 

The kinetic equations considered in phenomenological models of biochemical oscillations involve nonlinearities of a mixed linear-Hill type. The dynamical systems are typically finite dimensional, the number of degrees of freedom corresponding to the number of different substrates one takes into account. The use of Hill functions of the form 
$\chi_{y,n}(x)=x^n/(y^n+x^n)$ is motivated by phenomenological arguments and their ability to continuously interpolate between 0 and 1. The latter feature allows for a dynamical mechanism of switching, a property whose origin can be traced back to bifurcations of nonlinear maps and catastrophe theory. 

Of particular interest are theoretical studies of {\it circa-rhythms\/}. 
Let us recall that circa-rhythms are 
``classes of rhythms that are  capable of free-running in constant conditions with a period approximating that of the environmental cycle to which they are normally synchronized" 
\cite{Moore}. The examples of the rhythms are the circadian (24 hours), the circalunar (28 hours), or the circannual (365.25 days) ones. What makes circa-rhythms interesting from a formal point of view is the interplay between the phases where there exists an external forcing (light--darkness periods, say), and the rhythmicity that sustains even after the external driving is switched off (as in experiments in constant darkness). Formally, the two cases may correspond to nonlinearities involving (or not) an explicit time dependence of some parameters. 

To give an example, the model of circadian rhythms in {\it Neurospora\/} \cite{Gonze2000} involves four variables and two types of Hill functions. The rhythmicity occurs in the model even without an external driving, but the case with explicit time dependence of coefficients is treated as well. 
Similar constructions are given in \cite{Gonze2002a,Gonze2002b}. 

In the present paper we address the problem of circa-rhythmicity from the perspective of 
Eq. (\ref{fvN}). The main idea is to consider a composite Hamiltonian system that, as a whole, is conservative. There are two sets of degrees of freedom corresponding to its two subsystems. When we look at certain averages associated with one of the subsystems we find that their time evolution can be determined by an effective density matrix whose dynamics is given again by an equation of the form (\ref{fvN}), but now with explicitly time-dependent coefficients in the nonlinear function $f(\rho)$. Therefore, the interaction between the two subsystems effectively turns one of them into a pacemaker that drives the other subsystem and generates its evolution with characteristic bursting patterns exibiting ``day" and ``night" phases. 
As opposed to the phenomenological models found in the literature we do not ``force" switching by putting Hill functions by hand. We take a simple quadratic nonlinearity satisfying the ``no feedback, no nonlinearity" condition. The motivation for this concrete choice comes from its simplicity and experience  showing that the nonlinear behavior found here is generic and analogous to the one found for more complicated $f$'s (cf. the example of $f(\rho)=\rho^q-2\rho^{q-1}$ with arbitrary real $q$ analyzed in \cite{UCKL}). The solutions we obtain are also interesting in themseleves as first examples of solutions of (\ref{fvN}) exhibiting more than one self-swiching. 

The paper is organized as follows. We begin with an example of a nonlinear von Neumann equation (\ref{fvN}) for a composite system and time independent nonlinearity. We then find its particular solution that is reducible to a new von Neumann equation for a subsystem but with explicitly time-dependent nonlinearity. We then generalize the example to the class of time-dependent von Neumann equations and show how to construct their solutions. Next we explicitly solve an example and illustrate its rhythmic properties.

\section{Preliminaries: Dynamics in a subsystem}

As a simple example consider 
\be
i\dot\varrho
&=&
[\omega (\underbrace{b_1^{\dag}b_1-b_2^{\dag}b_2}_J)+\big(1+ X\big)a^{\dag}a, \varrho]
-
[X a^{\dag}a, f(\varrho)]
\ee
where $[b_k,b_k^{\dag}]=1=[a,a^{\dag}]$ and 
\be
X=b_1+b_1^{\dag}+b_2+b_2^{\dag}
\ee
For $\varrho^2=\varrho$ we find 
\be
i\dot\varrho
&=&
[\omega J+a^{\dag}a, \varrho]
\ee
describing three independent degrees of freedom: harmonic oscillation with unit frequency combined with rotation in a plane with frequency $\omega$. 
For $\varrho^2\neq \varrho$ rotation and oscillation get nonlinearly coupled. 

Let us now eliminate the rotation by switching to a rotating reference frame 
\be
\rho=e^{i\omega J t}\varrho e^{-i\omega J t}
\ee
where the density matrix satisfies
\be
i\dot\rho
&=&
[\big(1+ X(t)\big)a^{\dag}a, \rho]
-
[X(t) a^{\dag}a, f(\rho)].\label{fvN'}
\ee
Denoting $Y=-i(b_1+b_2^{\dag}-b_1^{\dag}-b_2)$ we find
\be
X(t)
&=&
e^{i\omega J t}X e^{-i\omega J t}
=
X\cos\omega t
+
Y\sin\omega t
\nonumber\\
Y(t)
&=&
e^{i\omega J t}Y e^{-i\omega J t}
=
Y\cos\omega t
-
X\sin\omega t\nonumber
\ee
and, for any $t$ and $t'$, 
\be
{[X(t),X(t')]}=[Y(t),Y(t')]=[X(t),Y(t')]=0.
\ee
Since $X$, $Y$, and $a^{\dag}a$ commute we can introduce their joint eigenvectors 
\be
X|x,y,n\rangle
&=&
x|x,y,n\rangle\\
Y|x,y,n\rangle
&=&
y|x,y,n\rangle\\
a^{\dag}a|x,y,n\rangle
&=&
n|x,y,n\rangle\\
X(t)|x,y,n\rangle
&=&
x(t)|x,y,n\rangle\\
Y(t)|x,y,n\rangle
&=&
y(t)|x,y,n\rangle\\
x(t)
&=&
x\cos\omega t
+
y\sin\omega t
\\
y(t)
&=&
y\cos\omega t
-
x\sin\omega t.
\ee
Now take any time-independent normalized vector 
$$
|\psi\rangle=\int dx dy \psi(x,y)|x,y\rangle
$$ 
and make the Ansatz $\rho(t)=|\psi\rangle\langle\psi|\otimes w(t)$ where $w$ is a density matrix acting only on the oscillator degress of freedom. Denoting $g(\rho)=\rho-f(\rho)
=|\psi\rangle\langle\psi|\otimes g(w)$ we rewrite (\ref{fvN'}) as 
\be
{}&{}&
|\psi\rangle\langle\psi|\otimes i \dot w
=
|\psi\rangle\langle\psi|\otimes [a^{\dag}a,w]
\label{gvN}\\
&{}&\pp=
+
X(t)|\psi\rangle\langle\psi|\otimes a^{\dag}a g(w)
-
|\psi\rangle\langle\psi|X(t)\otimes  g(w)a^{\dag}a\nonumber
\ee
Taking matrix elements of both sides of (\ref{gvN}) between arbitrary $\langle x,y|$ and 
$|x',y'\rangle$ we obtain 
\be
{}&{}&
\psi(x,y)\overline{\psi(x',y')} i \dot w
=
\psi(x,y)\overline{\psi(x',y')} [a^{\dag}a,w]
\label{gvN'}
\\
&{}&\pp=
+
\psi(x,y)\overline{\psi(x',y')}\Big( x(t) a^{\dag}a g(w)
-
x'(t)g(w)a^{\dag}a\Big).\nonumber
\ee
The Ansatz is internally consistent only if $\psi(x,y)=\sqrt{\delta(x-x_0)\delta(y-y_0)}$. The diagonal $x=x'=x_0$, $y=y'=y_0$ leads to
\be
i\dot w
&=&
[a^{\dag}a, w]
-
x_0(t) [a^{\dag}a, g(w)],\label{effective}
\ee
$x_0(t)=x_0\cos \omega t+y_0\sin \omega t$, 
which is an equation of the form (\ref{fvN}) but with $f$ replaced by 
\be
\tilde f(w)=w-x_0(t) g(w)=\big(1-x_0(t)\big)w+x_0(t)f(w).
\ee
If $f$ satisfies the ``no feedback, no nonlinearity" condition $f(\rho)=\rho$ for 
$\rho^2=\rho$, the same holds for the effective $\tilde f$. 

\section{Quantum pacemaker and feedback}

The pacemaker is an oscillatory system whose state can be modified by an external entraining agent, a {\it zeitgeber\/} \cite{Moore}. Our main interest in this paper is in the free-running oscillation of the pacemaker $x_0(t)$, and the circa-rhythms it induces in the absence of the zeitgeber. 

The next level is the coupling between the pacemaker, which defines the clock 
``mechanism", and the observable-level circa-rhythms which define the ``hands" of the clock. We assume a nonlinear feedback between the hands and the mechanism, but the dynamics is nondissipative in the sense that the energy of the hands averaged over a single cycle of the oscillation is constant. 

Motivated by the analysis of the previous section we concentrate on the following class of equations 
\be
i\dot w (t)=f_1(t)[H,w (t)]+f_2(t)[H,f\big(w (t)\big)]\label{vN2}
\ee
which define the state of the hands of the clock. 
The ``no-feedback, no-nonlinearity" condition reads 
$w (t)=f_1(t)w (t)+f_2(t)f\big(w (t)\big)$ whenever 
$w (t)^2=w (t)$. Now assume we know a solution $w _0(t)$ of (\ref{fvN}). 
Then 
\be
w (t)
&=&
e^{-iH\int_0^t f_1(x)dx}w _0\big(\textstyle{\int_0^t f_2(x)dx}\big)
e^{iH\int_0^t f_1(x)dx}
\ee
is a solution of (\ref{vN2}) as can be verified by a direct calculation. 
The whole problem of solving (\ref{vN2}) reduces to finding a solution of 
(\ref{fvN}), which can be performed by soliton techniques \cite{LC,UCKL}. 

Let us consider a simple but generic example where 
$f(w )=(1- s)w + s w ^2$ and 
\be
f_1(t)=1+\epsilon \cos\omega t ,\quad 
f_2(t)=-\epsilon \cos\omega t 
\ee
The parameter $ s$ allows us to compare situations where the driven dynamics is linear ($ s=0$) and purely nonlinear ($ s=1$), and for the two cases investigate the role of the $\epsilon$s. Varying $ s$ we can also investigate stability properties of the rhythms under fluctuations of the feedback. 

\section{Hands of the clock}

The hands of the clock are described by the Hamiltonian
$
H=a^{\dag}a=\sum_{n=0}^\infty n|n\rangle\langle n|
$ 
of a harmonic oscillator type. 
The frequency of the oscillator is equal to unity, meaning that this is a reference frequency for all the other frequencies found in the system. 
Modelling the hands by a harmonic oscillator is quite natural for obvious reasons. The quantum oscillator has in addition the appealing property of being delocalized. The oscillations occur at the level of probabilities in position space and what we call ``hands" is a kind of a center-of-mass quantity. This type of oscillation is what one expects in a biological system, since classical oscillators such as pendula or springs are too ``mechanical" to be of any relevance. Intuitively, in our model, the hands move at time $t$ to the region of space where the concentration of probability is the biggest. The probability density may be regarded as a measure of state of a certain spatially extended object. In the absence of a feedback between the hands and the pacemaker, the hands oscillate harmonically with their own internal frequency. We shall see later that the nonlinear coupling may practically suppress the internal oscillations of the hands in certain intervals of time. What will remain are the sudden bursts occuring, roughly, with the period of the pacemaker. 

Quantum harmonic oscillator is an infinite dimensional dynamical system and, hence, a solution of von Neumann equations may be characterized by an arbitrary number of parameters determining the initial state of the hands. 
There exists a simple trick allowing to construct infinitely-dimensional solutions on the basis of a single finite-dimensional one. The trick exploits equal spacing of eigenvalues of $H$. With our choice of units the eigenvalues are given simply by natural numbers. Let us divide them into sets containing 
$N$ elements: $\{0,1,\dots,N-1\}$, $\{N,N+1,\dots,N+N-1\}$, and so on. 
Each such subset corresponds to a block in the Hamiltonian, and each block can be represented by a $N\times N$ diagonal matrix of the form
\be
H_k=k \bm 1+{\rm diag}(0,1,\dots,N-1)=k \bm 1+H_0.
\ee
As a consequence, in each block we have to solve the same matrix equation since
a restriction $w _k$ of $w $ to the $k$-th subspace satisfies 
\be
i\dot w_k=[H_k,f( w _k)]=[H_0,f( w _k)].
\ee
The job can be reduced to finding a sufficiently general solution of a 
$N\times N$ problem. 
In each subspace we can take a different initial condition and a different normalization of trace. The whole infinite-dimensional solution will take the form of a direct sum
\be
 w (t)=\oplus_{k=0}^\infty p_k w _k(t,p_k),\label{direct}
\ee
$\sum_{k=0}^\infty p_k=1$. The $k$-th part depends on $p_k$ in a complicated way since the function $f( w )$ is not 1-homogeneous, i.e.  
$f(\lambda w )\neq \lambda f( w )$. The inhomogeneity implies that 
change of normalization simultaneously rescales time; the normalization of probability implies that a change of $p_k$ in a $k$-th subspace influences all the other subspaces by making their dynamics faster or slower. In this sense the solution, in spite of its simplicity, is not a simple direct sum of independent evolutions. 

In order to illustrate the possible effects we can use the solutions derived in  \cite{I} for the simplest nontrivial case involving self-switching, i.e. for 
$N=3$ and quadratic nonlinearity. Of course, since $N=3$, the Hamiltonian $H$ and the solution $ w $ possess matrix elements describing transitions 
$1\leftrightarrow 2$, $2\leftrightarrow 3$, $1\leftrightarrow 3$, between three different basis vectors. Although the observation is in itself trivial, we want to stress here the formal analogy to the so-called three-variable models of bursting oscillations in enzyme reactions with autocatalytic regulation 
\cite{DecrolyGoldbeter1982,DecrolyGoldbeter1987}. 

We select a subspace spanned by three subsequent vectors
$|k\rangle$, $|k+1\rangle$, $|k+2\rangle$. The family of interest is parametrized by $\alpha\in \bf R$ controlling the ``moment" and type of switching between bursts. The parameter naturally occurs at the level of the Darboux transformation, where it characterizes an initial condition for the solution of the Lax pair. 
The density matrices $ w _k(t)=\sum_{m,n=0}^2 w _{mn}
|k+m\rangle\langle k+n|$ are completely characterized by the $k$-independent matrix of time-dependent 
coefficients $ w _{mn}$. The reader may check by a straightforward substitution \cite{Math} that the matrix
\be
\left(
\begin{array}{ccc}
 w _{00} &   w _{01} &  w _{02} \\ 
 w _{10} &   w _{11} &  w _{12} \\
 w _{20} &   w _{21} &  w _{22} 
\end{array}
\right)
=
\frac{1}{15+\sqrt{5}}
\left(
\begin{array}{ccc}
5 & \xi(t) & \zeta(t) \\
\bar \xi(t) & 5+\sqrt{5} & \xi(t) \\
\bar \zeta(t) & \bar \xi(t) & 5
\end{array}
\right)\label{rhomatr}
\ee
with
\be 
\xi(t)
&=&
\frac{\left(2+3i-\sqrt{5}i\right)\sqrt{3+\sqrt{5}}\alpha}
{\sqrt{3}\big(e^{\gamma t}+\alpha^2
e^{-\gamma t}\big)}
e^{i\omega_0 t},
\nonumber\\
\zeta(t)
&=&
-
\frac{9e^{2\gamma t}+\left(1+4\sqrt{5}i\right)\alpha^2}
{3\big(e^{2\gamma t}+\alpha^2\big)}
e^{2i\omega_0 t}\nonumber
\ee
is indeed a normalized ($\Tr  w =1$) solution of the von Neumann equation 
\be
i\dot w =[H,(1- s) w + s w ^2],
\ee
$H={\rm diag}(0,1,2)$. 
The parameters are
$\omega_0=1-\frac{5+\sqrt{5}}{15+\sqrt{5}} s$, 
$\gamma=\frac{2}{15+\sqrt{5}} s$. 

Now let us rescale the trace. We do it in three steps. The modified density matrix 
\be
 w _1(t)=e^{i(1- s)Ht} w (t)e^{-i(1- s)Ht}
\ee
is a solution of 
$
i\dot  w _1(t)
=
[ s H, w _1(t)^2].
$
Therefore
\be
 w _2(t)=\Lambda e^{i(1- s)\Lambda Ht} w (\Lambda t)e^{-i(1- s)\Lambda Ht}
\ee
is also a solution of 
$
i\dot  w _2(t)
=
[H, s  w _2(t)^2]
$
and
\be
 w _3(t)=e^{-i(1- s)Ht} w _2(t)e^{i(1- s)Ht}
\ee
is a solution of 
\be
i\dot w _3(t)=[H,(1- s) w _3(t)+ s w _3(t)^2].
\ee
Performing these operations on our explicit solution we find 
\be
 w _3(t)
&=&
\frac{\Lambda }{15+\sqrt{5}}
\left(
\begin{array}{ccc}
5 & \xi_3(t) & \zeta_3(t) \\
\bar \xi_3(t) & 5+\sqrt{5} & \xi_3(t) \\
\bar \zeta_3(t) & \bar \xi_3(t) & 5
\end{array}
\right)
\ee
with
\be 
\xi_3(t)
&=&
\frac{\left(2+3i-\sqrt{5}i\right)\sqrt{3+\sqrt{5}}\alpha}
{\sqrt{3}\big(e^{\gamma \Lambda t}+\alpha^2
e^{-\gamma \Lambda t}\big)}
e^{i\big(\omega_0 \Lambda +(1- s)(1-\Lambda )\big)t},\nonumber\\
\zeta_3(t)
&=&
-
\frac{9e^{2\gamma \Lambda t}+\left(1+4\sqrt{5}i\right)\alpha^2}
{3\big(e^{2\gamma \Lambda t}+\alpha^2\big)}
e^{2i\big(\omega_0 \Lambda +(1- s)(1-\Lambda )\big)t},\nonumber
\ee
Now 
$
\int_0^tf_1(x)dx
=
t+\frac{\epsilon }{\omega}\sin\omega t $,
$\int_0^tf_2(x)dx
=
-\frac{\epsilon }{\omega}\sin\omega t $. 
We finally obtain the solution 
\be
 w (t)
&=&
\frac{\Lambda }{15+\sqrt{5}}
\left(
\begin{array}{ccc}
5 & \xi(t) & \zeta(t) \\
\bar \xi(t) & 5+\sqrt{5} & \xi(t) \\
\bar \zeta(t) & \bar \xi(t) & 5
\end{array}
\right)
\ee
with
\be 
\xi(t)
&=&
\frac{\left(2+3i-\sqrt{5}i\right)\sqrt{3+\sqrt{5}}\alpha}
{\sqrt{3}\big(e^{-\gamma \Lambda \frac{\epsilon }{\omega}\sin\omega t }+\alpha^2
e^{\gamma \Lambda \frac{\epsilon }{\omega}\sin\omega t }\big)}\nonumber\\
&\pp=&\times
e^{-i\big(\epsilon \omega_0 \Lambda +\epsilon (1- s)(1-\Lambda )-\epsilon \big)
\frac{\sin\omega t }{\omega}}
e^{it}
\nonumber\\
\zeta(t)
&=&
-
\frac{9e^{-2\gamma \Lambda \frac{\epsilon }{\omega}\sin\omega t }
+\left(1+4\sqrt{5}i\right)\alpha^2}
{3\big(e^{-2\gamma \Lambda \frac{\epsilon }{\omega}\sin\omega t }+\alpha^2\big)}\nonumber\\
&\pp=&\times
e^{-2i\big(\epsilon \omega_0 \Lambda +\epsilon (1- s)(1-\Lambda )-\epsilon \big)
\frac{\sin\omega t }{\omega}}
e^{2it}
\nonumber
\ee
of
\be
i\dot w (t)
&=&
\big(1+\epsilon \cos\omega t \big)[H, w (t)]\nonumber\\
&\pp=&-\epsilon \cos\omega t [H,(1- s) w (t) + s w (t)^2]
\ee
Since $\Tr H  w (t)=\Tr H  w (0)$, $\Tr H  w (t)^2=\Tr H  w (0)^2$, 
the internal energy of the system, averaged over one period $T$ of the pacemaker oscillation, is 
\be
E
=
\frac{1}{T}\int_{t}^{t+T}dt' \Tr H f\big( w (t')\big)
=
\Tr H  w (0),
\ee
and does not depend on $t$. In this sense the subsystem is conservative.  

If one does not integrate over the pacemaker period, one finds that the internal energy of the hands harmonically oscillates with the pacemaker frequency $\omega$. What is characteristic, however, the hands do not oscillate harmonically but behave as if they were accumulating energy during the phases of quiescence in order to suddenly release it in violent bursts. Moreover, the bursts are different during the two halves of the pacemeaker cycle, and thus resemble the day-night differences one finds in real organisms. 

All these properties are illustrated on explicit examples. 
We plot the dynamics of the hands $\bar q(t)=
\Tr \hat q w (t)=\int_{-\infty}^\infty dq \,q \langle q| w (t)|q\rangle$
as functions of time for different parameters characterizing the nonlinearity, and for different initial conditions. 

\section{Discussion}

The ability of plants and mammals to measure approximately 24-hour intervals of time is no doubt of an endogenous origin. Our current understanding of circadian rhythms leads to a notion of a pacemaker, an oscillating timing system whose role is to coordinate various complicated and different in shape rhythms of a biological system.

Splitting of a ``clock" into a ``pacemaker" and ``hands" occurs in our example quite naturally. The entire ``clock" is a conservative system so the rhythmic behavior we find is of endogenous origin. However, the system cannot be treated as isolated from external environment since the solution 
$\rho=P\otimes w$ is not a projector and, hence, involves nontrivial external correlations 
(cf. the analysis of this point given in \cite{I}). 

To make our analysis simple we have played with a concrete family of exact solutions of 
(\ref{fvN}). The family is characterized by several parameters related to the choice of the equation ($\omega$, $s$) and the initial condition for the dynamics ($x_0=-\epsilon$, 
$\alpha$). The solutions we have derived reveal an unknown aspect of nonlinear von Neumann equations --- the possibilty of evolution involving an infinite number of switching events.  

In spite of its simplicity, the quadratic nonlinearity generates circa-rhythmic behavior of surprising complexity and variety. Much more complicated types of behavior would have been found if one had used the freedom of higher dimensionality offered by (\ref{direct}). Moreover, the solution of the factorized form $P\otimes w$, with $P$ a projector, is very special and occurs only in cases the operators 
$X(t)$ and $X(t')$ commute for all $t$, $t'$. Had we replaced the Abelian group of rotations in a plane (with generator $J$) by something more complicated, we would not have been able to use the   Ansatz with factorization.

To end these remarks let us make it clear that we do not claim to have found a better description of nonlinear oscillations occuring in biochemical systems than the phenomenological approach of Goldbeter and his Brussels school. What we do claim, however, is that nonlinear von Neumann equations of the form (\ref{fvN}), in addition to their appealing general properties, imply a vast range of phenomena with analogies in biophysical systems. One should not be surprised if (\ref{fvN}) will one day play a fundamental role in life sciences.

\acknowledgments

Our work was supported by the KBN grant No. 5 P03B 040 20 (MC) and the Flemish Fund for Scientific Research (FWO project G.0335.02).

\begin{figure}\label{fig-5}
\includegraphics{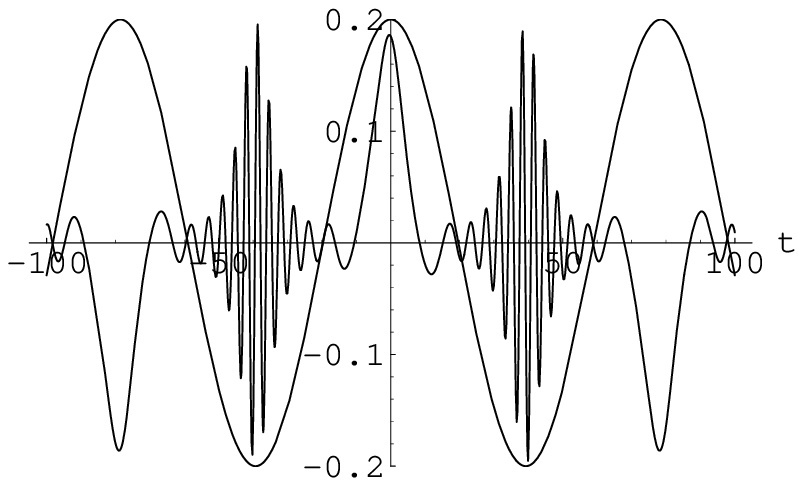}
\caption{Circa-rhythm of average position $\Tr \hat q w(t)$ for $\epsilon=2$, $\omega=0.08$, 
$\alpha=1$, $ s=-1.1$, $\Lambda=1$. The pacemaker sinusoidal oscillation is shown as a reference.}
\end{figure}
\begin{figure}\label{fig-7}
\includegraphics{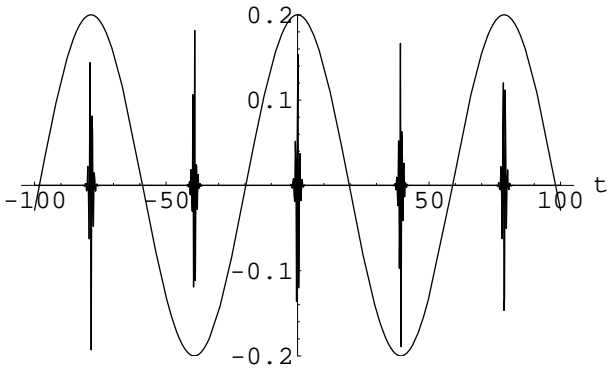}
\caption{The same parameters as in Fig.~1 but with $ s=-10$, i.e. for stronger nonlinearity.}
\end{figure}
\begin{figure}\label{fig-8}
\includegraphics{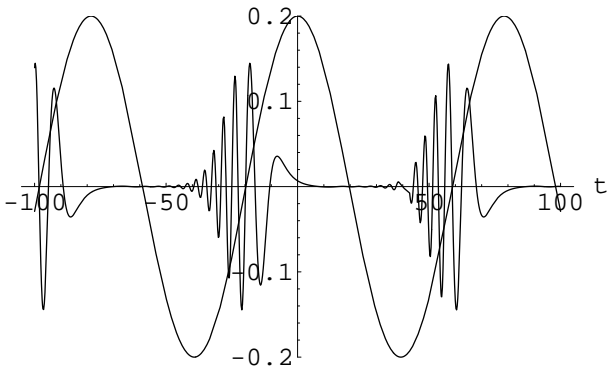}
\caption{The same parameters as in Fig.~1 but with $\alpha=e^{-4}$.}
\end{figure}
\begin{figure}
\includegraphics{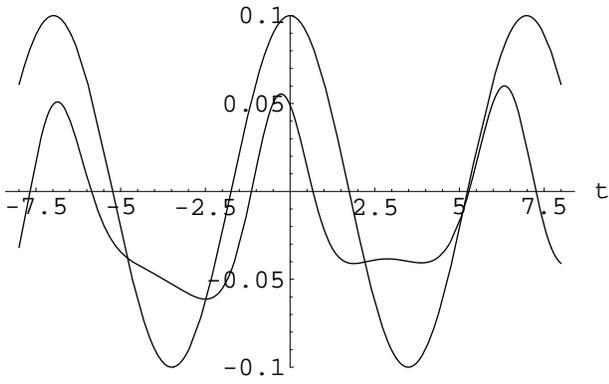}
\caption{$\Tr \hat q w(t)$ for $\epsilon=2$, $\omega=0.9$, 
$\alpha=e^2$, $ s=1$, $\Lambda=1$. Sinusoidal oscillation of the pacemaker shown for reference.}
\end{figure}
\begin{figure}
\includegraphics{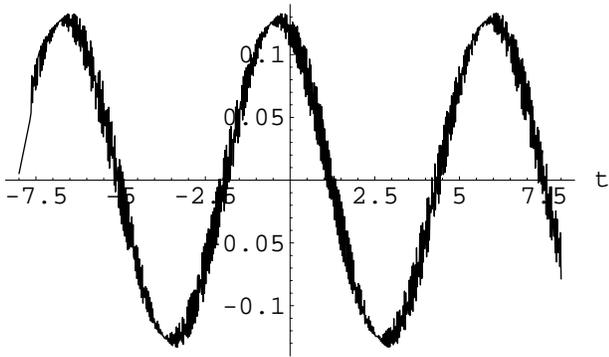}
\caption{$\Tr \hat q w(t)$ for $\epsilon=2$, $\omega=50$, 
$\alpha=e$, $ s=10$, $\Lambda=1$. The fast pacemaker oscillation occurs in a form of a 
``noise" and is not separately plotted.}
\end{figure}
\begin{figure}
\includegraphics{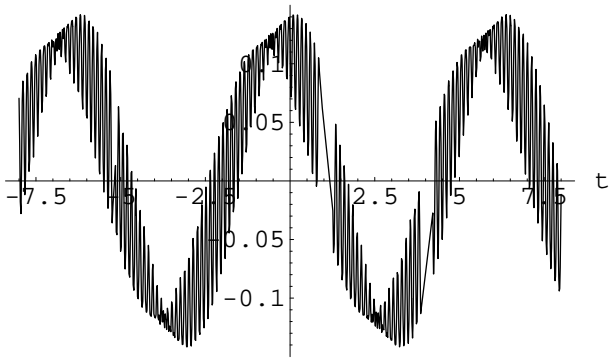}
\caption{Similar situation as in Fig.~5 but for $\epsilon=2$, $\omega=50$, 
$\alpha=e$, $ s=30$, $\Lambda=1$.}
\end{figure}
\begin{widetext}
\begin{figure}
\includegraphics{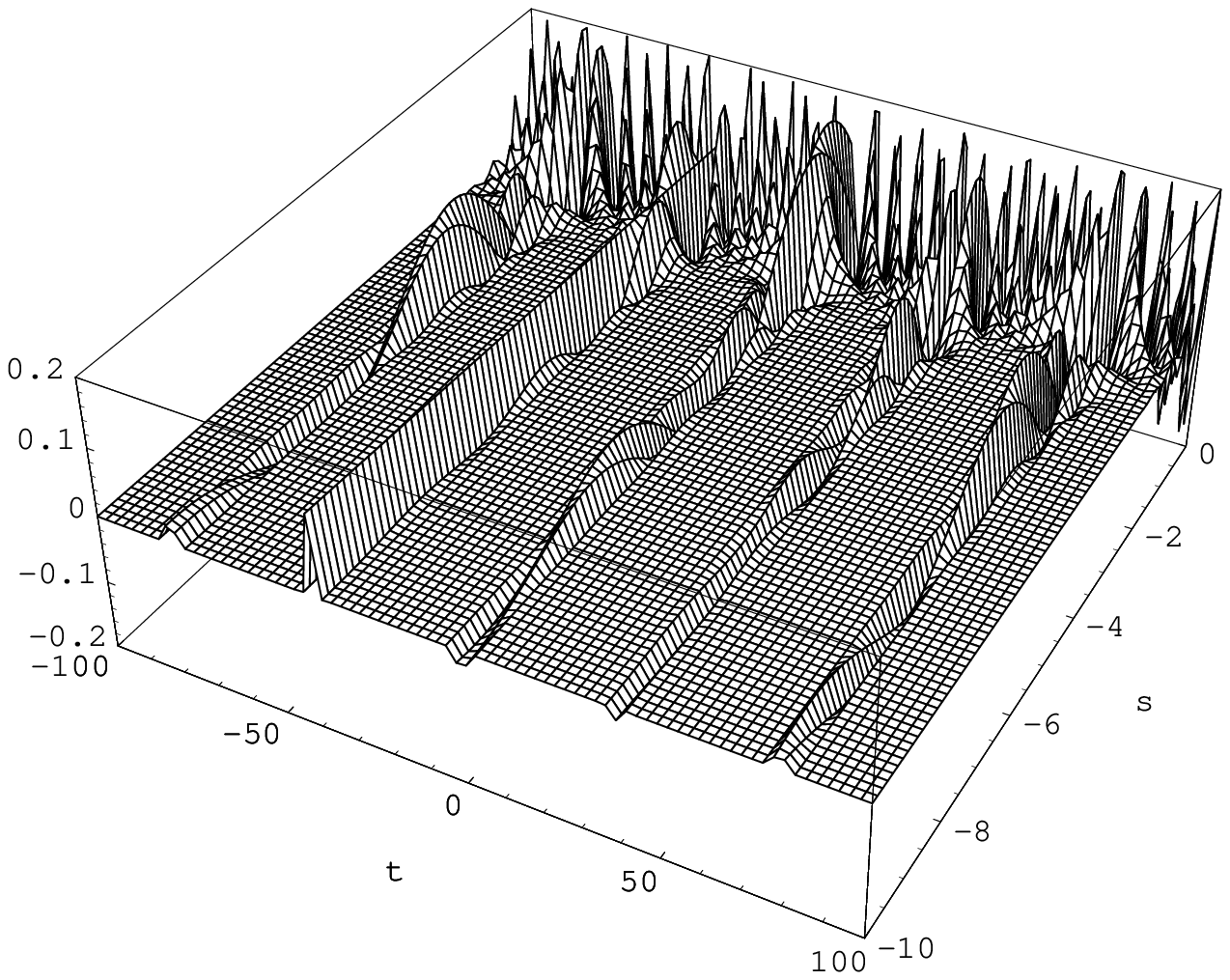}
\caption{Stability of circa-rhythmicity under changes of nonlinear coupling. Average position $\Tr \hat q w(t)$ for $\epsilon=2$, $\omega=0.08$, 
$\alpha=1$, $\Lambda=1$, and $-10\leq s\leq 0$. For $s$ sufficiently far from 0 the bursts do not qualitatively change with fluctuations of $s$. Fine details of the bursts are smeared out by coarse-graining of the plot. }
\end{figure}
\end{widetext}


\begin{references}
\bibitem{Math}All the examples discussed in the paper were obtained and tested by means of Mathematica 4.2.
\end{references}
\end{document}